\documentclass[aps,prl,preprint,groupedaddress]{revtex4-1}
\usepackage{graphicx}

\begin{document}


\title{The Changed Crystal Filed of $Gd^{3+}$ in Natural Zircon with Heat Treatment in Oxidizing and Reducing Atmosphere Monitor by ESR Spectroscopy}


\author{Araya Mungchamnankit$^1$, Suwimon Ruengsri$^{2,3}$}
\affiliation{$^1$Department of Physics, Faculty of Science, Rangsit University, Pathumthani, 12000, Thailand}
\affiliation{$^2$Department of Chemistry, Faculty of Science and Technology,
	Nakhon Pathom Rajabhat University, Nakhon Pathom, 73000, Thailand}
\affiliation{$^3$Center of Excellence in Glass Technology and Materials Science (CEGM),
	Nakhon Pathom Rajabhat University, 73000, Thailand
}


\date{\today}

\begin{abstract}
The main purpose of this research is dedicated to the ESR study of the impurity ions such as $Gd^{3+}$ in the natural zircon crystal because the presence of the impurity ions in the crystal usually causes color in crystal. The Zeeman interaction, the weak hyperfine interaction and the crystal field interaction due to the environments of the impurity ions will be calculated. Furthermore, the relation between the resonance magnetic field positions in the ESR spectra and the applied magnetic field directions will reflect the impurity ion site symmetry. The change in the crystal field parameters of heat-treated zircon at different atmospheres were related to the change in the color of zircon. 
\end{abstract}


\maketitle

\section{Introduction}
Zircon is unique among gemstones because of the remarkable range observed in its specific gravity and its refractivity \cite{IRECHE393,MUNGCHAMNANKIT2008479}. Zircon has been popular in Thailand, colorless zircon is well known for its properties including: brilliance and flashes of multicolored light, close to the properties of diamond. The source is often associated with corundum and good quality zircons are found in Thailand and Cambodia. Heat treatment is the important method for color enhancement of gemstones. Each type of gemstone has different conditions of heat treatment.  The improvement of color in gemstones depends on the atmosphere in the heating part of the furnace. Dark brown zircon, which was heated under an oxidizing atmosphere, became yellowish brown or colorless, while heated under a reducing atmosphere, it became light blue \cite{LA-ICP,ACHIWAWANICH20068646,doi:10.1142/S0217984901002646,KITTIAUCHAWAL2012706}.

In 2006, we began to study the impurity ions in zircon, using laser ablation inductively coupled plasma mass spectrometry (LA-ICP-MS) and electron spin resonance (ESR) spectroscopy. The main purpose of this research was dedicated to the ESR study of the impurity ions such as Gd$^{3+}$ in the natural zircon crystal because the presence of the impurity ions in the natural crystal usually causes the color in crystal. In the following year, we experimented to heat treating zircon at 900 $^\circ$C in oxygen and argon atmosphere with calculated crystal field parameters. This experiment affects the change in the second degree crystal field parameter ($B_2^0$). The gadolinium ion was in the presence of a large axial crystalline electric field generated from its surrounding oxygen ions within the zircon lattice \cite{IRECHE393,MUNGCHAMNANKIT2008479}. In 2012, we learned more about the effect of CO$_2$ atmosphere on color changing in zircon and found that the optimized temperature for heat-treating zircon was 900 $^\circ$C for 6 hours. The color was changed from dark brown to greenish blue with more clarity in the zircon crystal. The crystal filed was not calculated.

According to previous work, we know the mechanism of heat-treating zircon in the oxidizing and reducing atmosphere had two different shades. In oxidizing atmosphere, zircon turns dark brown to light yellowish brown. Zircon changes to greenish blue and yellowish blue when heat treated in a reducing atmosphere. In the term of crystal field, temperature and heat treatment will have the same effect that this research shows the change of crystal field in spin Hamiltonian parameters by ESR spectroscopy. 
\section{Experimental Setup}
\begin{figure}[ht]
\includegraphics[scale=0.9]{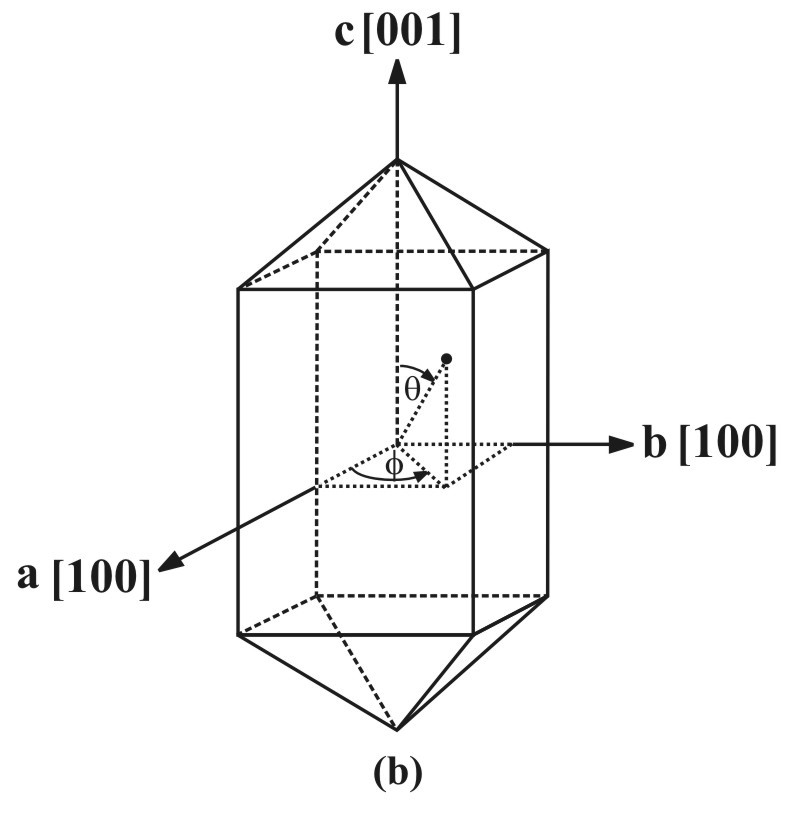}%
\caption{\label{fig:crystal}The rotation angle $\theta$ and $\phi$ [2].}
\end{figure}

The natural zircons were obtained from Cambodia. The zircon crystal samples were cleaned up using H$_2$SO$_4$ and distilled water to remove impurities on their surfaces. The samples of dimension about 2$\times$2$\times$5 mm$^3$ and known c-axis were selected for ESR measurements. After that, the samples were separately heated at 900 $^\circ$C in oxygen and CO$_2$ atmosphere. ESR measurements were performed at room temperature (293 K) in the microwave range of X-band ($\sim$9.86 GHz) with the usual 100 kHz field modulation on a Bruker ELEXYS E500 CW ESR spectrometer. The sample was mounted on a quartz sample holder in which it could be rotated about its axis. The ESR spectra were measured with the c-axis [001] parallel and perpendicular to the applied magnetic field. The magnetic field was varied from 0 to 650 mT. The ESR spectra were recorded every 15$^\circ$ of the rotation angle, $\phi$ about c-axis and the rotation angle, $\theta$ about x or y-axis from 0 to 180$^\circ$ as shown in Fig. \ref{fig:crystal}. The variations of the magnetic field positions of the rotation angles $\theta$ and $\phi$ were observed to confirm the impurity ion site symmetry in the zircon crystal. The spin Hamiltonian parameters and energy level diagrams were calculated from ESR spectra.
\section{Results and Discussion}
\begin{figure}[ht]
	\includegraphics[scale=0.55]{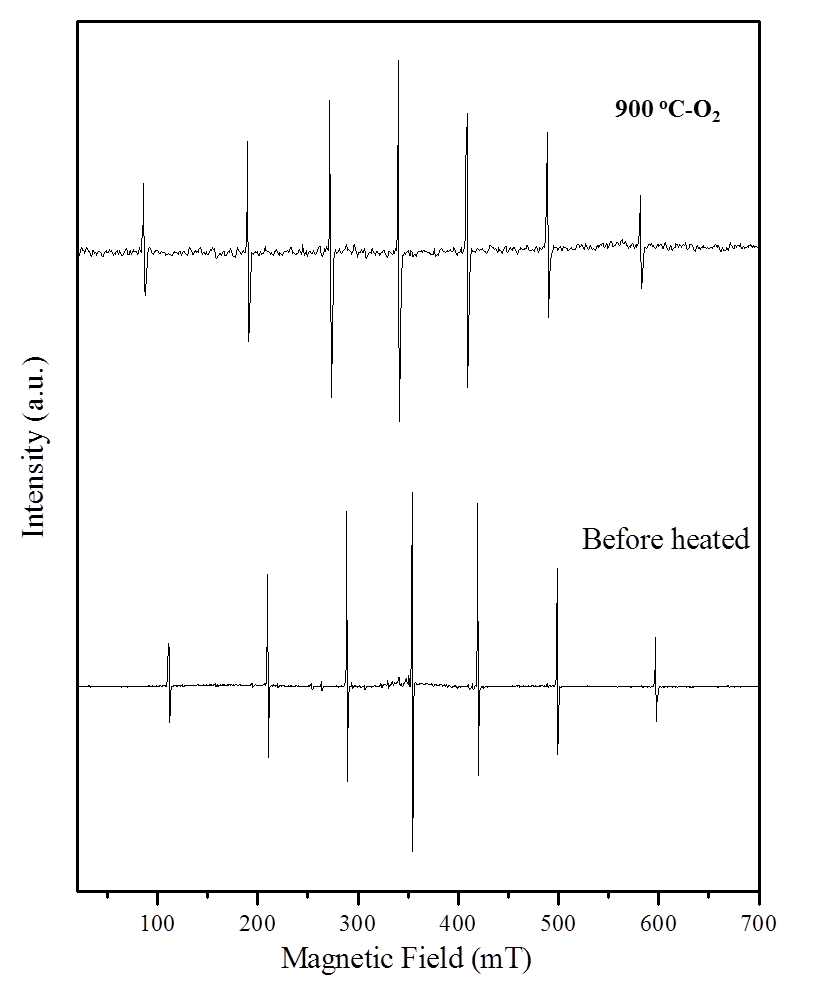}%
	\caption{\label{fig:graph1}ESR spectra of zircon before and after heat treatment in oxygen atmospheres at 900 $^\circ$C with the applied magnetic field parallel to the c-axis.}
\end{figure}

\begin{figure}[ht]
	\includegraphics[scale=0.55]{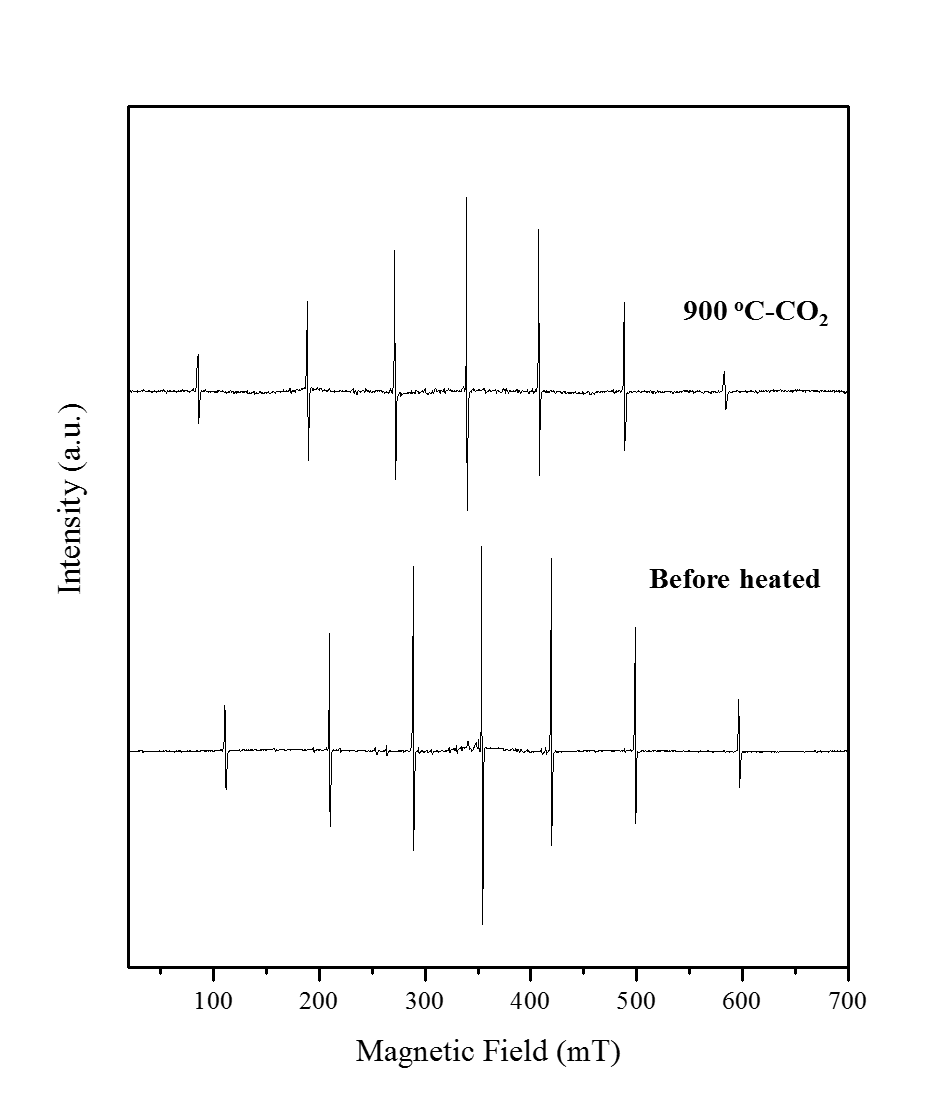}%
	\caption{\label{fig:graph2}ESR spectra of zircon before and after heat treatment in CO$_2$ atmospheres at 900 $^\circ$C with the applied magnetic field parallel to the c-axis.}
\end{figure}

The ESR spectra of zircon before and after heat treatment in oxygen and CO$_2$ atmosphere were observed in X-band when the applied magnetic fields are parallel to the c-axis [001] at room temperature (298 K) as shown in Fig. \ref{fig:graph1} and \ref{fig:graph2}. The seven resonance absorption peaks of Zeeman interaction are obtained due to a trace of the Gd$^{3+}$ present in the natural zircon. Since the electron configuration of Gd$^{3+}$ is [Xe]4f$^7$, hence the spin quantum number, S = 7/2 and the ground state of Gd$^{3+}$ is $^8$S$_{7/2}$. The state $^8$S$_{7/2}$ splits into 2S+1 = 8 states, then the seven allowed transitions according to the selection rule, $\Delta$M$_S$ = $\pm$1, are obtained \cite{MUNGCHAMNANKIT2008479,LA-ICP,ACHIWAWANICH20068646,doi:10.1142/S0217984901002646,KITTIAUCHAWAL2012706}. The ESR spectra can be described by a spin Hamiltonian incorporating with Zeeman interaction, hyperfine structure and crystal field operators \cite{Tennant2004,doi:10.1063/1.1671333,doi:10.1063/1.1677080} are given by equations:\
\begin{eqnarray}
	H & = & \beta\mathbf{S}\cdot\mathbf{g}\cdot\mathbf{B}+B_2^0 O_2^0+B_4^0 O_4^0+B_4^4 O_4^4+B_6^0 O_6^0+B_6^4 O_6^4
	\label{eq:hyperfine}
\end{eqnarray}
The first term corresponds to the Zeeman interaction arises from the interaction between electron spin angular momentum and external magnetic where $\beta$, $\mathbf{S}$, $\mathbf{g}$ and $\mathbf{B}$ are the Bohr magneton, the spin operators, gyromagnetic tensor and magnetic field respectively.

\begin{figure}[t]
	\includegraphics[scale=0.5]{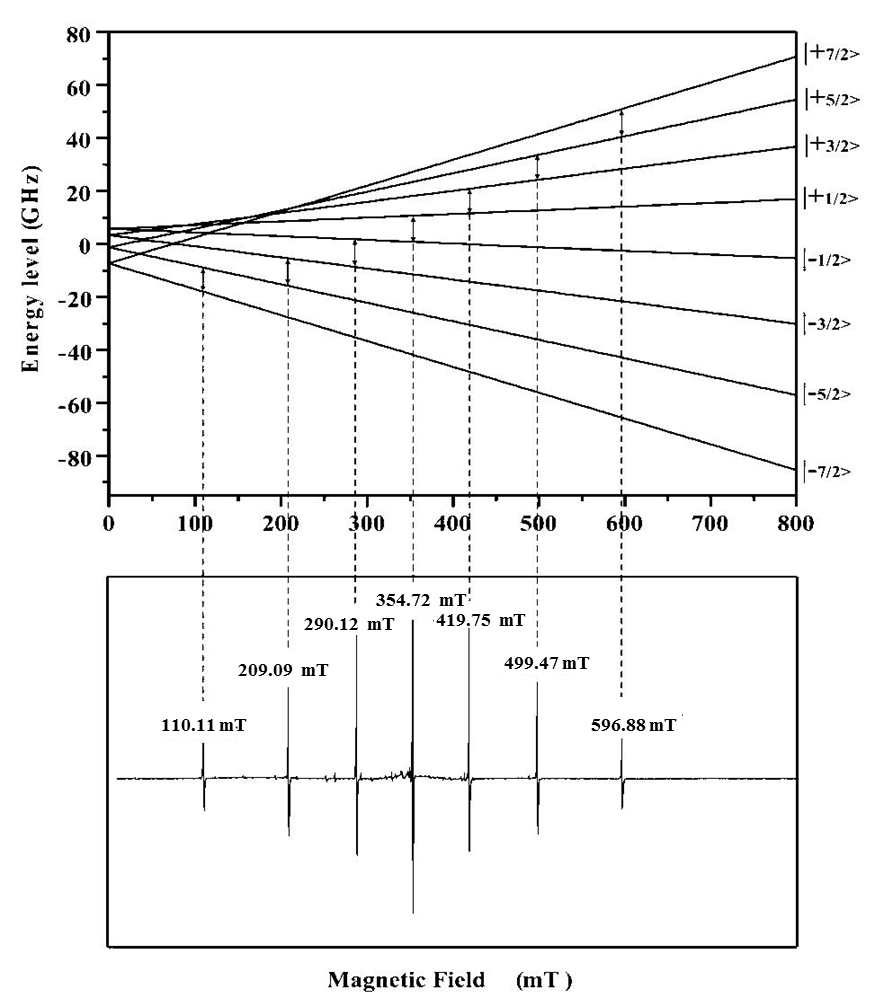}%
	\caption{\label{fig:graph3}Energy level diagram of Gd$^{3+}$ in unheated zircon as a function of the applied magnetic field parallel to the c-axis (above), and the transitions corresponding to the measured peaks of ESR spectrum (below).}
\end{figure}

The second term is the crystal field term which arises from the crystal field potential generated from the surroundings of the paramagnetic ion in zircon. This term depends on the local symmetry of paramagnetic ion site and the electronic configuration of ion. It is the sum of the spin angular momentum operators called Stevens’ operators or the equivalent operators ($O_2^0$, $O_4^0$, $O_4^4$, $O_6^0$ and $O_6^4$) with their coefficients ($B_2^0$, $B_4^0$, $B_4^4$, $B_6^0$ and $B_6^4$) as the crystal field parameters. The parameters in equation (\ref{eq:hyperfine}) were calculated from the resonance magnetic field positions in the ESR spectra and the obtained parameters are shown in Table \ref{tab:result}. The principal spin Hamiltonian parameters   were determined by measuring the magnetic field positions of the ESR spectra both in the applied magnetic fields parallel and perpendicular to the c-axis.
\begin{table}[h]
	\caption{\label{tab:result}Spin Hamiltonian parameters of Gd$^{3+}$ in zircon before and after heat treatment in oxygen and argon atmosphere.
	}
\begin{ruledtabular}
	\begin{tabular}{ccccc}
		Parameters & \multicolumn{2}{c}{$O_2$ atmosphere} & \multicolumn{2}{c}{$CO_2$ atmosphere}\\
		\cline{2-3} \cline{4-5}
		& before heated & after heated & before heated & after heated \\
		\hline
		$g_{xx}=g_{yy}$ & 1.9914 & 1.9912 & 1.9914 & 1.9910 \\
		$g_{zz}$ & 1.9919 & 1.9920 & 1.9919 & 1.9918 \\
		$B_2^0$ & 376.45 & 373.24 &	369.73 & 378.56 \\
		$B_4^0$ & 7.02 & 7.80 &	8.05 & 7.92 \\
		$B_4^4$ & 78.29 & 76.91	& 76.01 & 78.64 \\
		$B_6^0$ & 0.65 & 0.75 &	0.68 & 0.59 \\
		$B_6^4$ & 0.12 & 0.12 &	0.12 & 0.12 \\
	\end{tabular}
\end{ruledtabular}
\end{table}
The constant $g_{zz}$, $B_2^0$, $B_4^0$ and $B_6^0$ could be accurately determined by perturbation theory when the magnetic field is parallel to the c-axis. The off-diagonal matrix elements $O_4^4$ and $O_6^4$ are considerably smaller than the diagonal ones for sufficiently high resonance frequency. Since the corresponding spin Hamiltonian parameters $B_4^4$ and $B_6^4$ are usually small, the contributions arising from these terms can be neglected for the orientation $\mathbf{B}\parallel$ c. Under these approximations, the seven allowed transitions ($\Delta$M$_S$ = $\pm$1) occur at the magnetic field positions given by the following first order expressions \cite{doi:10.1063/1.440259}:
\begin{eqnarray}
	\pm\frac{7}{2}\leftrightarrow\pm\frac{5}{2}:\,\mathbf{B} & = & \frac{1}{g\beta}[h\nu\pm\left(-6B_2^0-20B_4^0-6B_6^0\right)]\nonumber\\
	\pm\frac{5}{2}\leftrightarrow\pm\frac{3}{2}:\,\mathbf{B} & = & \frac{1}{g\beta}[h\nu\pm\left(-4B_2^0+10B_4^0+14B_6^0\right)]\\
	\pm\frac{3}{2}\leftrightarrow\pm\frac{1}{2}:\,\mathbf{B} & = & \frac{1}{g\beta}[h\nu\pm\left(-6B_2^0-20B_4^0-6B_6^0\right)]\nonumber\\
	+\frac{1}{2}\leftrightarrow-\frac{1}{2}:\,\mathbf{B} & = & \frac{h\nu}{g\beta},\nonumber
\end{eqnarray}
where $\nu$ is the magnetic frequency.  The ground state energy level of the Gd$^{3+}$ in the natural zircon that is without a magnetic field, was split into four doublets, denoted by $|\pm 7/2\rangle $, $|\pm 5/2\rangle $, $|\pm 3/2\rangle $  and $|\pm 1/2\rangle $, due to crystal field or zero field splitting. When the magnetic field was applied to the zircon, the magnitude of splitting becomes larger due to Zeeman interaction. The transition between the two energy levels follow the selection rule $\Delta$M$_S$ = $\pm$1. By using the parameters in Table \ref{tab:result}, we can simulate the energy level diagrams (EPR-NMR program) and the results for unheated zircon are shown in Fig. \ref{fig:graph3}. It was found that the transition energy between the two simulated energy levels of Zeeman splitting line in Fig. \ref{fig:graph3} are nearly the same as the absorbed microwave energy obtained from the ESR experiments. From the ESR results, The Zeeman interaction, the weak hyperfine interaction, and the crystal field interaction due to the environments of the impurity ions were calculated. Particularly, the changes in the crystal field parameters of heat-treated zircon at different atmospheres were investigated and these were related to the change in color of zircon.
\section{Conclusions}
The ESR spectra of zircon crystal before and after heat-treated at 900 $^\circ$C in oxygen and CO$_2$ atmosphere with the applied magnetic field parallel and perpendicular to the c-axis [001] directions were generated from the microwave energy absorptions due to the transitions between the spin states of gadolinium ions (Gd$^{3+}$, S = 7/2), which were the impurity ions in the natural zircon crystal. These were confirmed by the clearly resolved fine structure of the gadolinium ion.
The second degree crystal field parameter ($B_2^0$) is much larger than the other crystal field parameters. This result reveals that the gadolinium ion is in the presence of a large axial crystalline electric field generated from its surrounding oxygen ions with in the zircon lattice. This electric field of tetragonal symmetry has caused the ESR spectra with the applied magnetic field direction. The change in the crystal field parameters of heat-treated zircon at different atmospheres were related to the change in color of zircon.
\section{Acknowledgments}
\begin{acknowledgments}
	The authors wish to thanks Department of Chemistry, Faculty of Science, Mahidol University for ESR measurement. This work is financially supported by Research Institute of Rangsit University, Thailand
\end{acknowledgments}
\bibliography{ref}

\end{document}